\documentclass[letterpaper,12pt]{article}
\usepackage{osajnl2}
\usepackage{subfig}
\usepackage{grffile}
\usepackage{float}
\graphicspath{%
  {../figs/illustrations/}%
  {../figs/input/}%
  {../figs/results/}%
}

\begin{document} \title{Phase
  retrieval combined with digital holography} \author{Eliyahu
  Osherovich$^{*}$, Michael Zibulevsky, and Irad Yavneh}
\address{Computer Science Department, Technion ---
  Israel Institute of 
  Technology, \\ Haifa, 32000, Israel} \address{$^*$Corresponding
  author: oeli@cs.technion.ac.il}

\begin{abstract}
We present a new method for real- and complex-valued image
reconstruction from two intensity measurements made in the Fourier
plane: the Fourier magnitude of the unknown image, and the
intensity of the interference pattern arising from superimposition
of the original signal with a reference beam. This approach can
provide significant advantages in digital holography since it
poses less stringent requirements on the reference beam. In
particular, it does not require spatial separation between the
sought signal and the reference beam. Moreover, the reference beam
need not be known precisely, and in fact, may contain severe
errors, without leading to a deterioration in the reconstruction
quality. Numerical simulations are presented to demonstrate the
speed and quality of reconstruction.
\end{abstract}

\ocis{100.5070, 100.3175, 090.1995.}

\maketitle{}

\section{Introduction}
\label{sec:introduction}

The reconstruction of an image from the magnitude of its Fourier
transform, also known as the phase retrieval problem, is of
paramount importance in a variety of modern imaging techniques.
Notable relevant disciplines include astronomy, crystallography,
and microscopy. Hence, the images in question may vary from remote
stars and galaxies to microscopic objects such as single
cells\cite{thibault06reconstruction}, viruses
\cite{song08quantitative}, and even atomic structure of nano-scale
objects like, for example, carbon nano-tubes\cite{zuo03atomic}.
One of the first successful numerical methods for phase retrieval
was developed by Gerschberg and Saxton (GS) with application to
electron microscopy \cite{gerchberg72practical}. Basically, their
method can be viewed as a complex-valued image reconstruction from
two intensity measurements: the magnitude of the image itself, and
the magnitude of its Fourier transform. Starting with the GS
method, Fienup later developed a new reconstruction method called
the Hybrid-Input-Output (HIO) algorithm, that requires only one
measurement, namely, the Fourier transform magnitude of the sought
signal \cite{fienup82phase}. To
the best of our knowledge, HIO is currently the most widely used
computational method for phase retrieval. However, there are
several situations where this method fails. Most often HIO is
unable to reconstruct complex-valued objects whose support (the
locus of the object's non-vanishing parts) is not known precisely
\cite{fienup87reconstruction}. Moreover, in certain situations HIO
may stagnate without converging to the correct image
\cite{fienup86phase-retrieval,wackerman91use}. Other
significant drawbacks of the HIO algorithm are relatively slow
convergence and sensitivity to measurement errors
\cite{osherovich08signal,osherovich09fast}. In some cases it may
require thousands of iterations before a moderately sized image is
reconstructed to high accuracy
\cite{osherovich09fast,osherovich11approximate}. Several attempts
have been made to address the above problems by using continuous
optimization methods such as Newton-type algorithms.
Unfortunately, these methods fail when applied to the phase
retrieval problem \cite{nieto-vesperinas86study}. A simple
heuristic explanation of this failure was given
in\cite{osherovich11approximate}. Alongside this explanation it
was demonstrated that a rough Fourier phase estimate may be
sufficient to allow successful application of Newton-type
optimization methods to the phase retrieval problem
\cite{osherovich09fast,osherovich11approximate}. In the present
work we remain within the framework of introducing additional
information into the phase retrieval problem so as to develop
faster and more robust methods of reconstruction. Specifically, we
consider the situation where two intensity measurements in the
Fourier domain are available. One is the Fourier magnitude of the
sought image, as in classical phase retrieval, and the second is
the intensity pattern resulting from the interference of the
original signal with a known reference beam also measured in the
Fourier plane. Although either one of these measurements can, in
theory, be sufficient for successful reconstruction of the unknown
image, our method provides significant advantages over such
reconstructions. For example, comparing with reconstruction from
the Fourier magnitude alone by HIO, our method gives a much faster
speed and better quality in case of noisy measurements (see
\cite{osherovich11approximate,osherovich09fast,osherovich08signal}).
Furthermore, unlike classical holography methods, our algorithm
does not require any special design of the reference beam. Finally
and most importantly, very good reconstruction quality is obtained
even when the reference beam contains severe errors.

The rest of the paper is organized as follows. In
Section~\ref{sec:basic-reconstr-algor} we present the details of
our basic reconstruction method which is developed for the
situation where the reference beam is known precisely. The
relation between our setup and digital holography is discussed in
Section~\ref{sec:relation-holography}.
Section~\ref{sec:reconstr-meth-impr} is devoted to further
development of our reconstruction method to accommodate possible
errors in the reference beam. Numerical simulation results are
presented in Section~\ref{sec:results}. Finally, a short
discussion and concluding remarks are given in
Section~\ref{sec:conclusions}.

\section{Basic reconstruction algorithm}
\label{sec:basic-reconstr-algor} Let us start with the notation
used throughout the paper. The unknown two-dimensional signal that
we wish to reconstruct is represented by the complex-valued
function $U(x,y)=|U(x,y)|\exp[j\varphi(x,y)]$, where
$\varphi(x,y)$ designates the object phase, and $j$ is the
imaginary unit: $j=\sqrt{-1}$. To address the phase of a
complex-valued number we use the angle notation: $\angle(U(x,y))
\equiv \varphi(x,y)$. Our measurements are done in the Fourier
plane $(\xi,\eta)$, hence the transformation that $U$ undergoes
when transforming from the $(x,y)$ plane to the $(\xi,\eta)$ plane
is simply the unitary Fourier transform
\begin{equation}
  \label{eq:1}
  \hat{U}(\xi,\eta) = \mathcal{F}\{U(x,y)\}\ .
\end{equation}
Strictly speaking, the actual transformation is a bit more
complicated as it includes some constant multipliers and scale
factors \cite{goodman04introduction}.  However, these are
immaterial for our discussion.  Hereinafter, we shall adopt the
convention that a pair of symbols like $X$ and $\hat{X}$ denote a
signal in the $(x,y)$ plane (also known as the object domain) and
its counterpart in the $(\xi,\eta)$ plane (also referred to as the
Fourier domain), respectively. For the sake of brevity, we may
omit the location designator $(x,y)$  or $(\xi,\eta)$ and use $X$
or $\hat{X}$ when the entire signal is considered.

The main purpose of our work is to develop a robust reconstruction
method that can tolerate severe errors in the reference beam. To
this end we use the reference beam only for \textit{estimating}
the Fourier phase of the sought image. Once a rough phase estimate
is available we can use the method of phase retrieval with
approximately known Fourier phase that was developed in
\cite{osherovich11approximate,osherovich09fast,osherovich08signal}.
This method was demonstrated to have fast convergence properties
and good quality of reconstruction from noisy measurements (see
\cite{osherovich11approximate,osherovich09fast,osherovich08signal}
for details and comparison with HIO).

\begin{figure}
  \centering
  \includegraphics{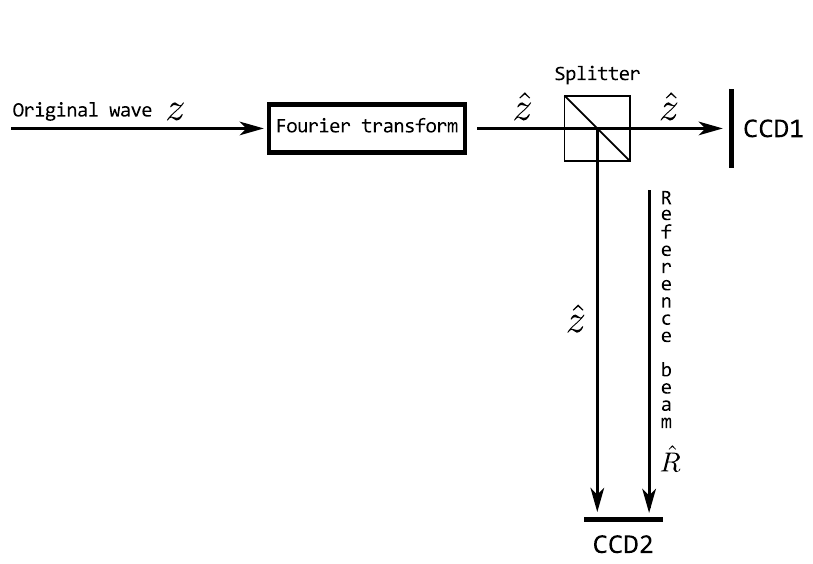}
  \caption{Schematic representation of the experiment.}
  \label{fig:experiment-schematic}
\end{figure}

The two measurements available at our disposal are used as follows.
$I_{1}$ provides the Fourier magnitude of the sought image via the
simple relationship between the two:
\begin{equation}
  \label{eq:2}
  I_{1}(\xi,\eta) = |\hat{U}(\xi,\eta)|^{2}\ .
\end{equation}
The second measurement reads
\begin{equation}
  \label{eq:3}
  I_{2}(\xi,\eta) =|\hat{U}(\xi,\eta) + \hat{R}(\xi,\eta)|^{2}\ ,
\end{equation}
where $\hat{R}(\xi, \eta)$ denotes a known reference beam that is
used to obtain the Fourier phase estimate as described below. One
possible schematic setup that provides these measurements is shown
in Fig.~\ref{fig:experiment-schematic}. Note that $\hat{R}$ is not
necessarily a Fourier transform of some physical signal $R$ in the
object domain. This means that $\hat{R}$ can be formed directly in
the Fourier plane without forming first $R$ and applying then an
optical Fourier transform to obtain $\hat{R}$.  Nevertheless,
there exists a mathematical inverse
\begin{equation}
  \label{eq:4}
  R(x,y) = \mathcal{F}^{-1}\{\hat{R}(\xi,\eta)\}\ ,
\end{equation}
whose properties, such as extent, magnitude, etc., can be
considered. The only requirement of $\hat{R}$ is that it must not
vanish in the region of our measurements. This is an important
point that provides an advantage to our method over the classical
holography techniques. We shall elaborate more on this in
Section~\ref{sec:relation-holography}.

\begin{figure}
  \centering
  \subfloat[]{
    \label{fig:projection-fourier}
    \includegraphics[width=0.4\textwidth]{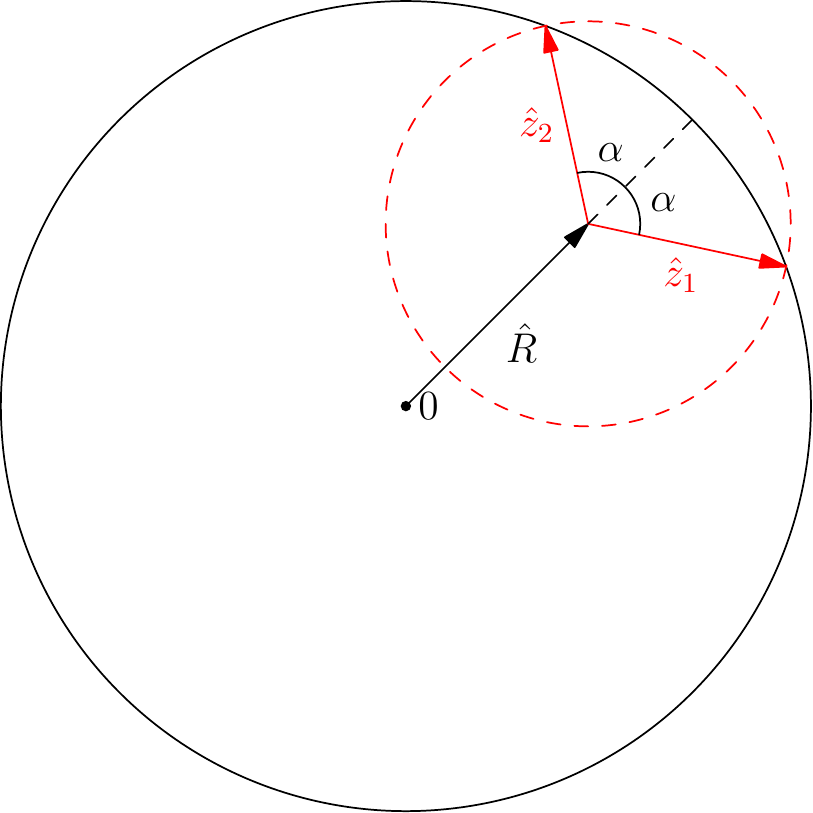}
  }
  \qquad{}
  \subfloat[]{
    \label{fig:convexrelaxation}
    \includegraphics[width=0.4\textwidth]{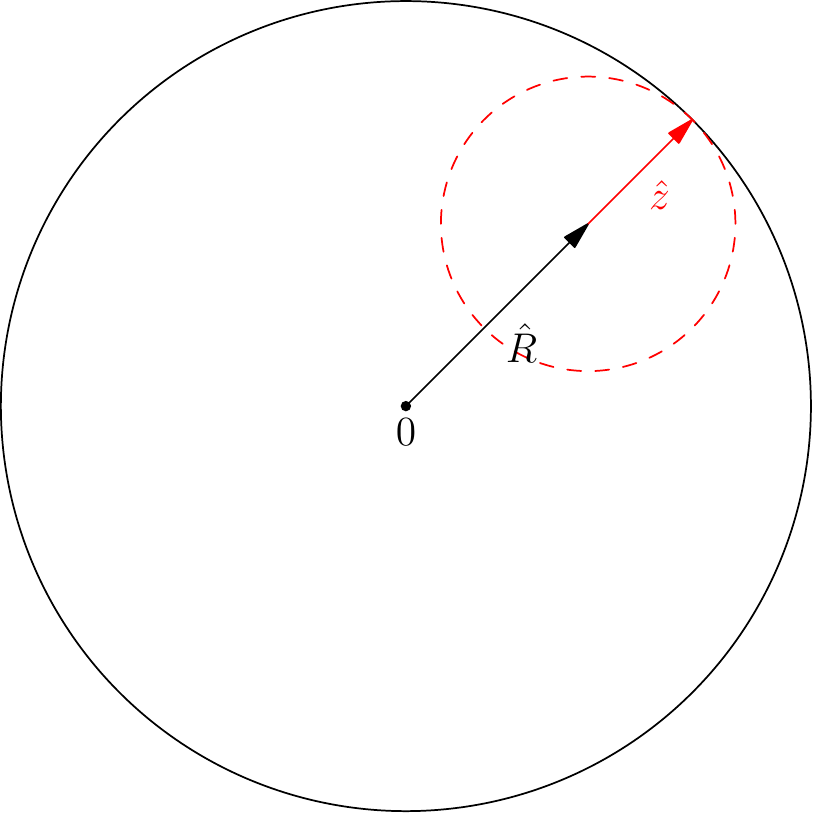}
  }
  \caption{Given a reference beam (black) whose magnitude and phase
    are known, and an unknown signal of known magnitude (dotted
    circle radius), one can try to find the phase of the unknown
    signal by measuring the magnitude of the sum (solid circle
    radius). Evidently, in most cases there are two possible
    solutions (a). However, in certain cases, the is only one
    solution (b).}
  \label{fig:holography-possible-situations}
\end{figure}

Let us now describe  how $\hat{U}$'s phase information is extracted
from $I_{2}$, and, more importantly, how it is used in our
reconstruction method. Consider the two signals:
\begin{equation}
  \label{eq:8}
  \hat{U}(\xi,\eta) = |\hat{U}(\xi,\eta)|\exp[j\phi(\xi,\eta)]\ ,
\end{equation}
and
\begin{equation}
  \label{eq:9}
  \hat{R}(\xi,\eta) = |\hat{R}(\xi,\eta)|\exp[j\psi(\xi,\eta)]\ .
\end{equation}
The intensity pattern of their interference can be written as:
\begin{eqnarray}
  \label{eq:10}
  I_{2}(\xi,\eta)
  & = &|\hat{U}(\xi,\eta) + \hat{R}(\xi,\eta)|^{2}\\
  & = &|\hat{U}(\xi,\eta)|^{2} + |\hat{R}(\xi,\eta)|^{2}
  + \hat{U}^{*}(\xi,\eta)\hat{R}(\xi,\eta)
  + \hat{U}(\xi,\eta)\hat{R}^{*}(\xi,\eta)\label{eq:13}\\
  & = & |\hat{U}(\xi,\eta)|^{2} + |\hat{R}(\xi,\eta)|^{2}
  +
  2|\hat{U}(\xi,\eta)|\,|\hat{R}(\xi,\eta)|\cos[\phi(\xi,\eta)-\psi(\xi,\eta)]
  \label{eq:7}
  \ .
\end{eqnarray}
From this formula we can easily extract the difference between the
unknown phase $\phi(\xi,\eta)$ and the known phase
$\psi(\xi,\eta)$:
\begin{equation}
  \label{eq:11}
  \cos[\phi(\xi,\eta)-\psi(\xi,\eta)] =
  \frac
  {I_{2}(\xi,\eta) -|\hat{U}(\xi,\eta)|^{2} - |\hat{R}(\xi,\eta)|^{2}}
  {2|\hat{U}(\xi,\eta)|\,|\hat{R}(\xi,\eta)|} \ .
\end{equation}
Which gives us:
\begin{equation}
  \label{eq:12}
  \phi(\xi,\eta) = \psi(\xi,\eta) \pm \alpha(\xi,\eta)\ ,
\end{equation}
where
\begin{equation}
  \label{eq:18}
  \alpha(\xi,\eta) = \arccos
  \left[
    \frac
    {I_{2}(\xi,\eta) -|\hat{U}(\xi,\eta)|^{2} - |\hat{R}(\xi,\eta)|^{2}}
    {2|\hat{U}(\xi,\eta)|\,|\hat{R}(\xi,\eta)|}
  \right]\ .
\end{equation}
This expression is well defined as $|\hat{R}(\xi,\eta)|$ is
assumed to be non-zero everywhere in the region of interest and
the places where $|\hat{U}(\xi,\eta)|=0$ can be simply excluded
from our consideration as there is nothing to be recovered since
their phase has no influence. We assume that $\pm\alpha$, that is,
the difference between the phases $\phi$ and $\psi$, lies within
the interval $[-\pi,\pi]$, hence, no phase unwrapping is
necessary. The phase $\phi(\xi,\eta)$ can assume either one
(rarely) or two possible values at every location. The two
possible situations are shown in
Fig.~\ref{fig:holography-possible-situations}. Hence, if the
intensity $I_{2}(\xi,\eta)$ is sampled at $N$ points there are
generally $2^{N}$ possible solutions $\hat{U}(\xi,\eta)$ and,
consequently, the same number of possible reconstructions
$U(x,y)$. (Here we consider the worst case scenario where all
sampled values give rise to two solutions.) To guarantee a unique
(and meaningful) reconstruction we must use additional information
about the sought signal $U(x,y)$. In the phase retrieval problem
it is usually assumed that $U(x,y)$ has limited support, namely,
part of the image is occupied by zeros.  In practice, it is
usually assumed that, in each direction, half (or more) pixels of
$U(x,y)$ are zeros. To capture this information in the Fourier
domain one should ``over-sample'' by a factor of two (or more) in
each dimension. Hence, if the known (not necessarily tight)
support area of $U$ is $n\times n$ pixels, then in the Fourier
plane it must be sampled with a sensor of size $2n\times 2n$
pixels. Such ``over-sampling'' usually guarantees unique (up to
trivial transformations: shifts, constant phase factor, and axis
reversal) reconstruction in the case of the classical phase
retrieval problem, where only $|\hat{U}|$ is available
\cite{hayes82reconstruction}. It is not known whether this
two-fold oversampling is absolutely necessary in our case where
two measurements are available. However, our experiments indicate
that for a general complex-valued signal it still seems to be
necessary to over-sample by a factor of two. Hence, the
reconstruction problem reads: find $U(x,y)$ such that $|\hat{U}|$
is known, $\angle(\hat{U})=\psi\pm\alpha$, and $U(x_{O},y_{O}) =
0$. Here, $\alpha$ is the known phase difference between $\hat{U}$
and $\hat{R}$ as defined by Equations~(\ref{eq:12})
and~(\ref{eq:18}); $(x_{O},y_{O})$ denotes the known off-support
area in the $(x,y)$ plane.

The problem is combinatorial in nature, and many different methods
can be applied to find a solution. Our method is based on
replacing the equality $\angle(\hat{U})=\psi\pm\alpha$ with the
less strict inequality
\begin{equation}
  \label{eq:5}
  \psi - \alpha \leq \angle(\hat{U}) \leq \psi + \alpha\ .
\end{equation}
By this relaxation we reduce the original problem into the phase
retrieval problem with approximately known phase.  As mentioned
earlier, for this situation there exists an efficient
reconstruction method based on quasi-Newton
optimization\cite{osherovich08signal,osherovich11approximate,osherovich09fast}.
The method solves the nonlinear non-convex optimization problem
defined by
\begin{equation}
  \label{eq:6}
  \left\{
    \begin{array}{rl}
      \displaystyle\min_{Z} & \||\mathcal{F}\{Z\}| -  |\hat{U}|\|^{2} \\
      \mathrm{s.\,t.} & \psi - \alpha \leq \angle(\mathcal{F}\{Z\}) \leq
      \psi+\alpha \, , \\
      & Z(x_{O}, y_{O}) = 0 \, ,
    \end{array}
  \right.
\end{equation}
by further relaxing the problem so as to perform the optimization
over a convex set (see \cite{osherovich09fast,osherovich08signal}
for details). Note that the solution of the above minimization
problem $Z$ is not guaranteed to be equal to $U$. Due to the
relaxation we performed, the phase of $\hat{Z}$ is allowed to
assume the continuum of values in the interval $[\psi-\alpha,
\psi+\alpha]$ instead of the two discrete values $\psi\pm\alpha$.
However, due to the uniqueness of the phase retrieval problem, the
phase of $Z$ may differ from the phase of $U$ only by a constant.
That is, $Z(x,y)=U(x,y)\exp[jc]$ for some scalar $c$
(see~\cite{hayes82reconstruction} for details). This does not pose
problems, as only the relative phase distribution inside the
support area of $U(x,y)$ is usually of interest. Moreover, in the
case where the absolute phase is required,  $c$ can be recovered
by adding a post-processing step that solves the one-dimensional
optimization problem:
\begin{equation}
  \label{eq:19}
  \displaystyle\min_{c} \||\hat{Z}\exp[jc] + \hat{R}| -
  I_{2}^{1/2}\|^{2}\ .
\end{equation}
Note that we intentionally do not add a penalty term like
$\||\hat{Z} + \hat{R}| - I_{2}^{1/2}\|^{2}$ into the main
minimization scheme as defined by Equation~(\ref{eq:6}).  Adding
such a term would introduce a strong connection between $Z$ and
the reference beam $\hat{R}$. This connection will inevitably
deteriorate the quality of reconstruction when the reference beam
$\hat{R}$ contains errors.

Reconstruction speed is very fast as is evident from our
experiments (see Fig.~\ref{fig:we-reconstruction-speed}).
Moreover, it can be further accelerated, as our experiments
indicate that more aggressive oversampling (zero padding in the
object domain) results in faster convergence in terms of the
number of iterations. In fact, use of a special reference beam
can, in theory, result in a trivial non-iterative reconstruction
in a way similar to holography. However, such special reference
beams may not be easily realizable in physical systems and the
quality of the reconstructed signal is strongly influenced by the
quality of the reference beam. We discuss this setup in the next
section and compare its sensitivity to possible errors in
$\hat{R}$ against our method in Section~\ref{sec:results}.

\section{Relation to holography} \label{sec:relation-holography}

Our method was initially developed for the phase retrieval
problem. However, the use of interference patterns creates a
strong connection with holography. Therefore, it may be pertinent
to discuss the advantages our method provides over the classical
holographic reconstruction. Note that our discussion is limited to
basic holography only and no attempt is made to cover all possible
setups and techniques that can be used in digital holography. We
nonetheless believe that this novel approach can compete with or
improve upon existing algorithms used in digital holography.

\begin{figure}
  \centering
  \subfloat[]{
    \label{fig:holography-real-square}
    \includegraphics{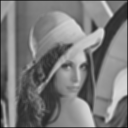}
  }
  \qquad{}
  \subfloat[]{
    \label{fig:holography-complexSmooth-square}
    \includegraphics{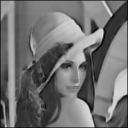}
  }
  \qquad{}
  \subfloat[]{
    \label{fig:holography-complexRandom-square}
    \includegraphics{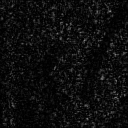}
  }
  \caption{Image (intensity) reconstructed by the holographic
    technique using a small square as the reference beam: (a) the
    image is real-valued, (b) image phase varies slowly, (c)
    image phase is random (varies rapidly).}
  \label{fig:holography-reconstruction-intensity}
\end{figure}

\begin{figure}
  \centering
  \subfloat[]{
    \label{fig:we-quality-real}
    \includegraphics[height=0.25\textheight]{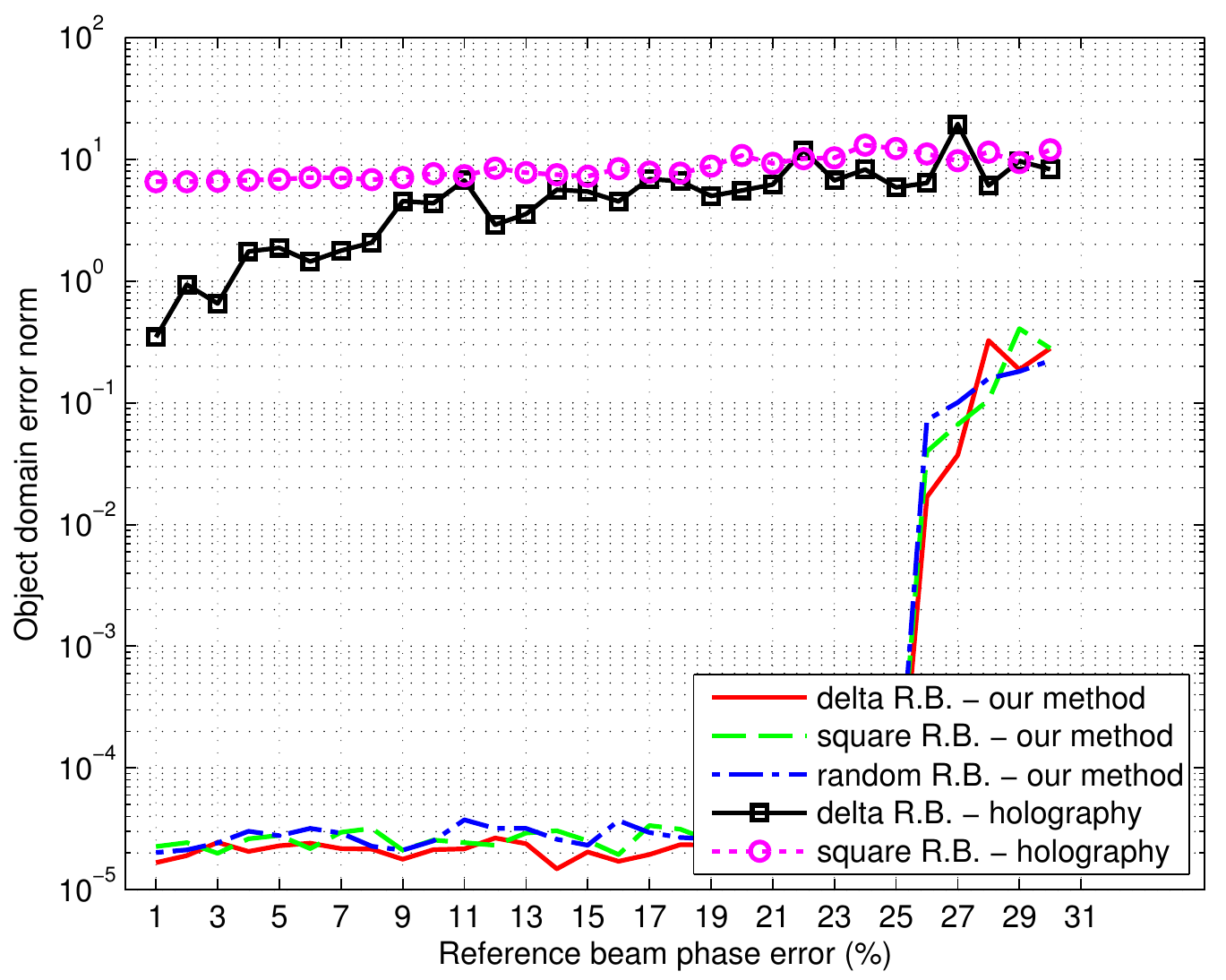}
  }
  \subfloat[]{
    \label{fig:we-quality-complexSmooth}
    \includegraphics[height=0.25\textheight]{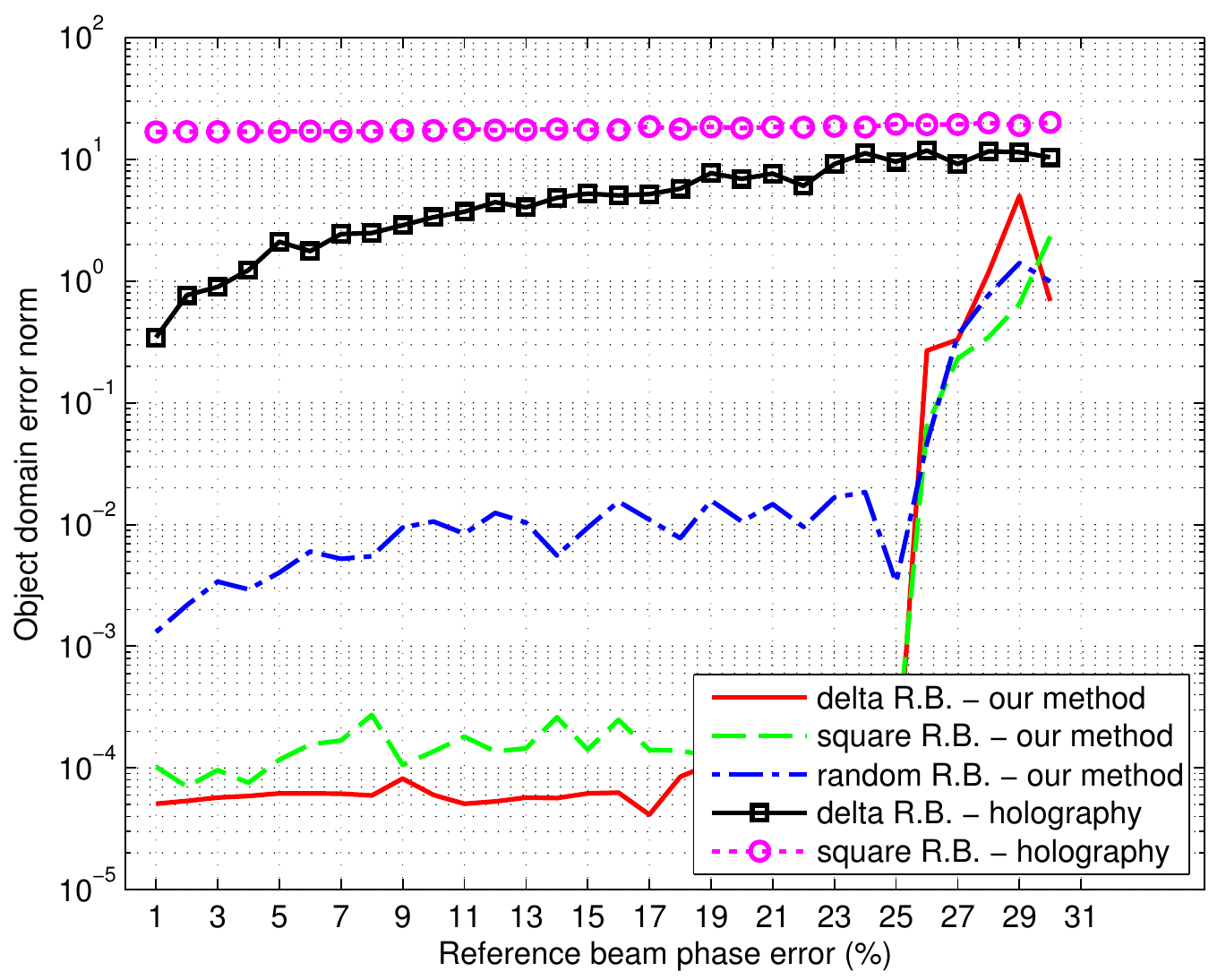}
  }
  \qquad{}
  \subfloat[]{
    \label{fig:we-quality-complexRandom}
    \includegraphics[height=0.25\textheight]{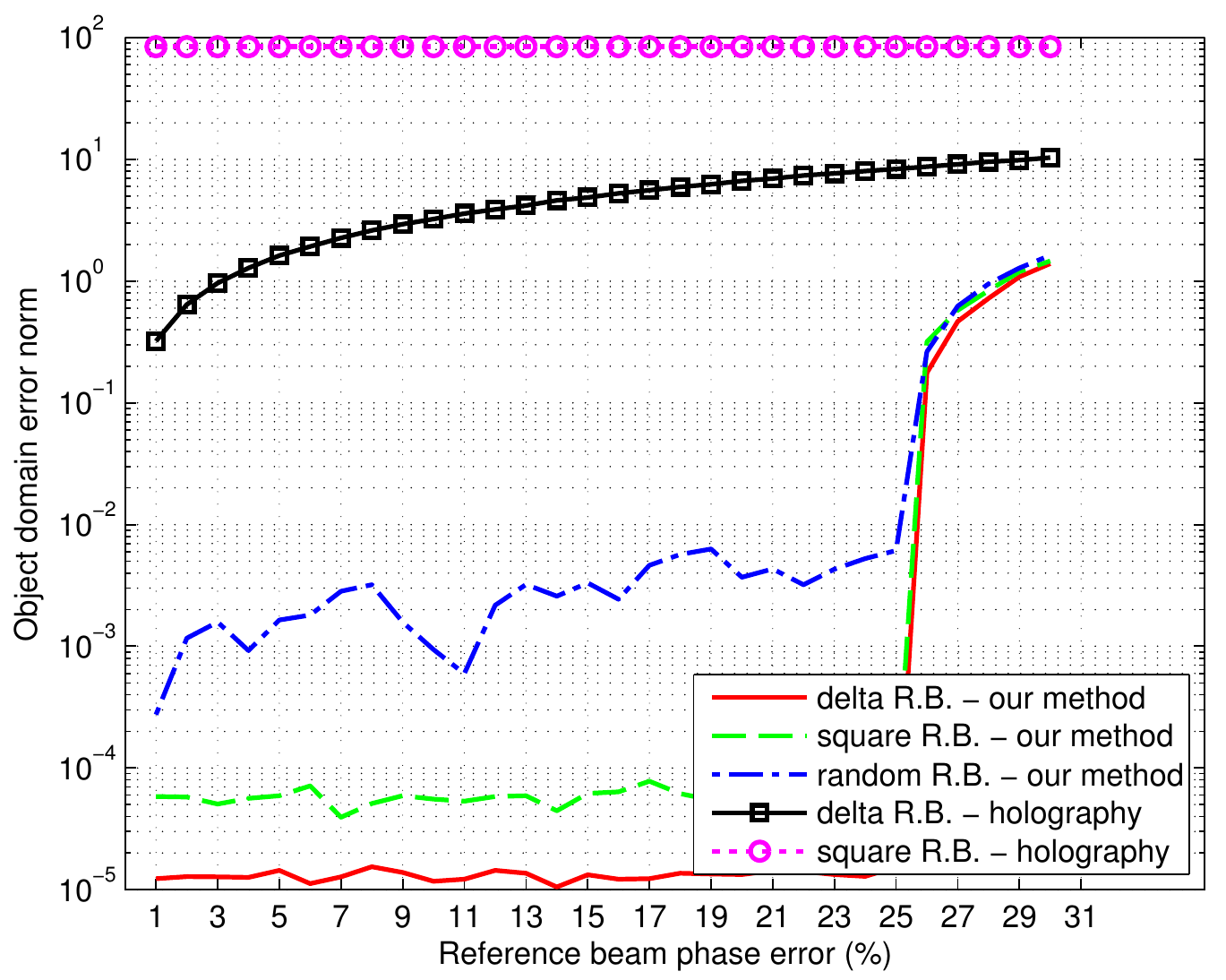}
  }
  \caption{Object domain error ($\|Z-U\|$) vs. phase error in the
    reference beam: (a) real valued image,
    (b) image phase is smooth, (c) image phase is random.}
  \label{fig:objectdomain-error}
\end{figure}

\begin{figure}
  \centering
  \subfloat[]{
    \label{fig:we-qualitycorrected-real}
    \includegraphics[height=0.25\textheight]{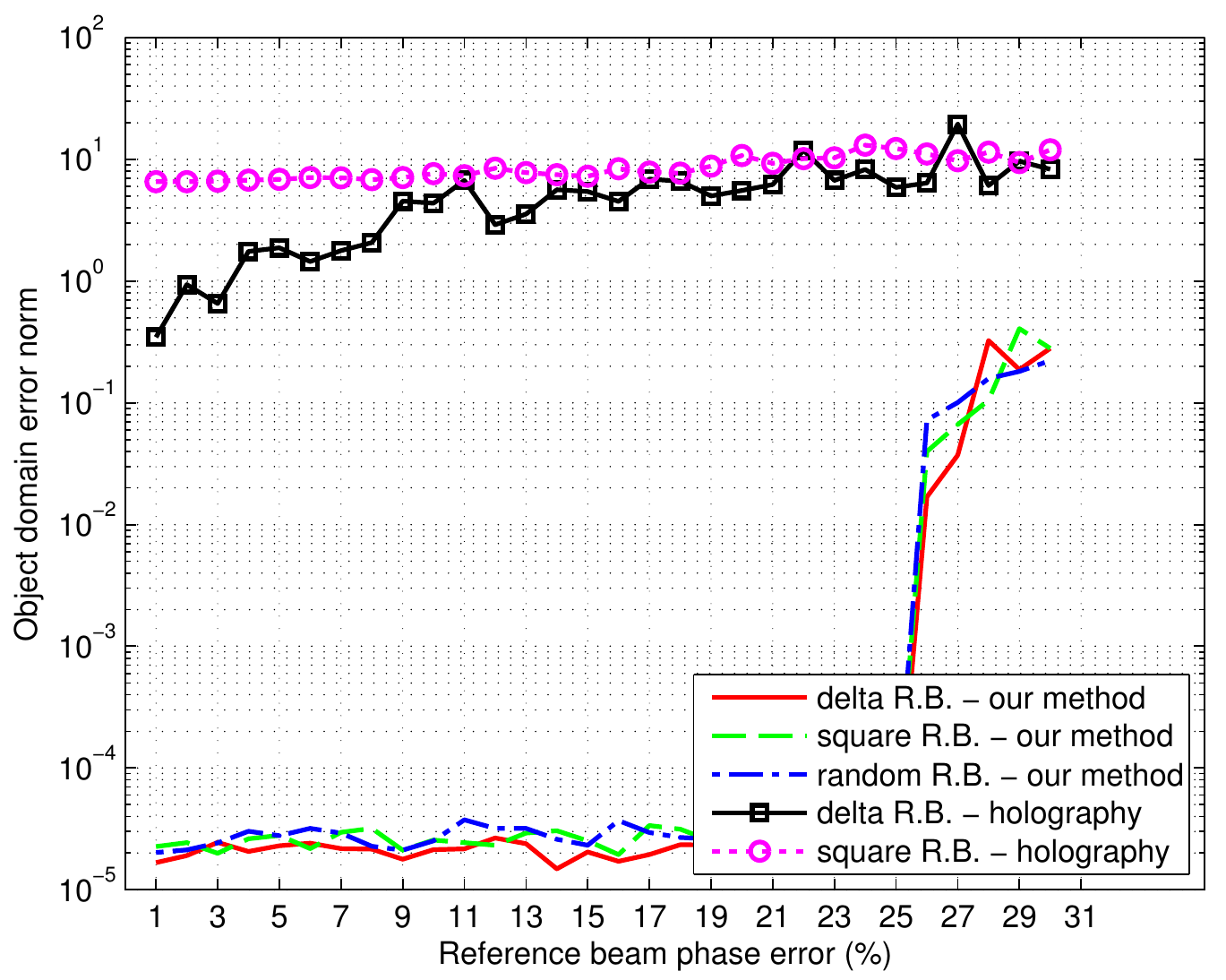}
  }
  \subfloat[]{
    \label{fig:we-qualitycorrected-complexSmooth}
    \includegraphics[height=0.25\textheight]{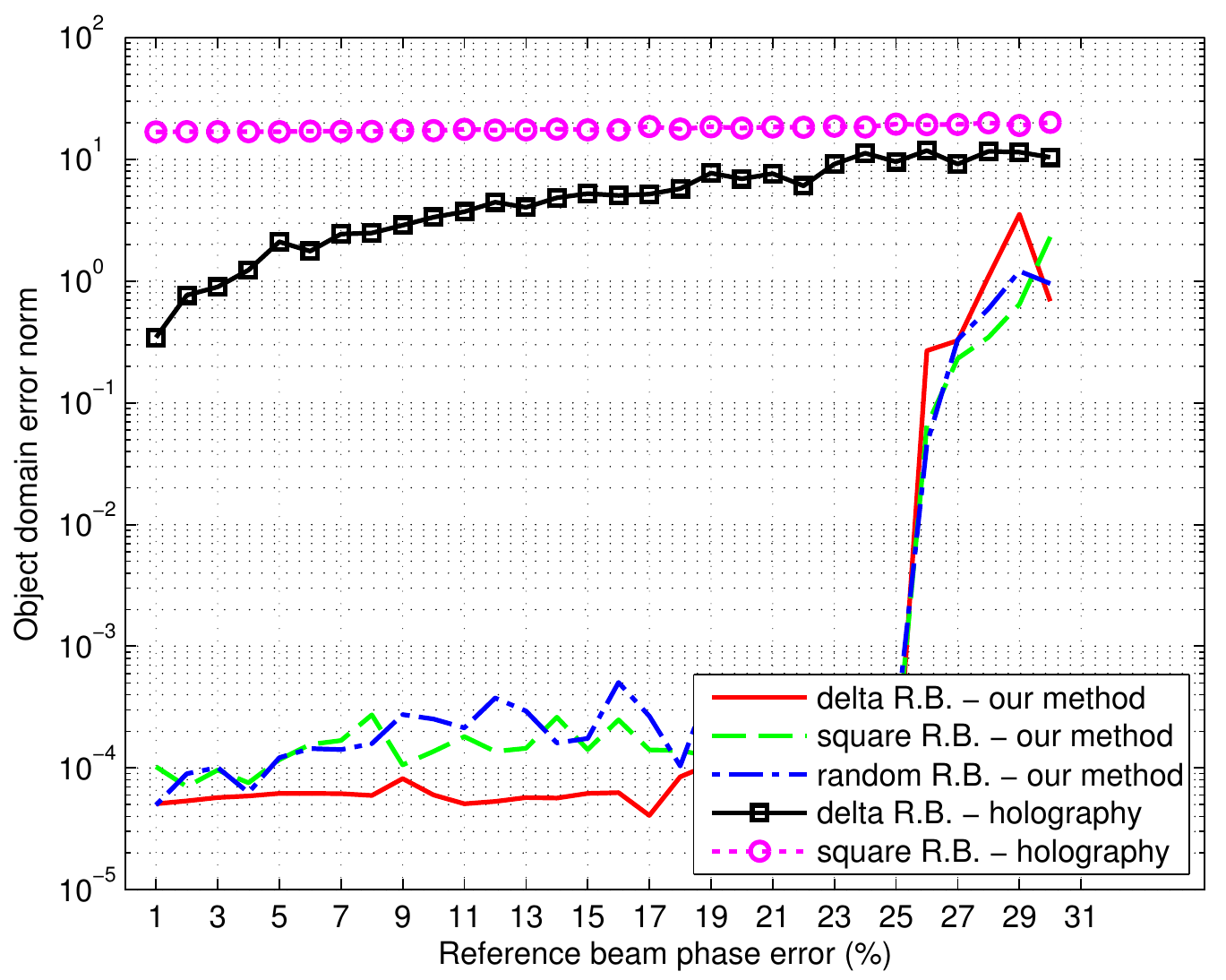}
  }
  \qquad{}
  \subfloat[]{
    \label{fig:we-qualitycorrected-complexRandom}
    \includegraphics[height=0.25\textheight]{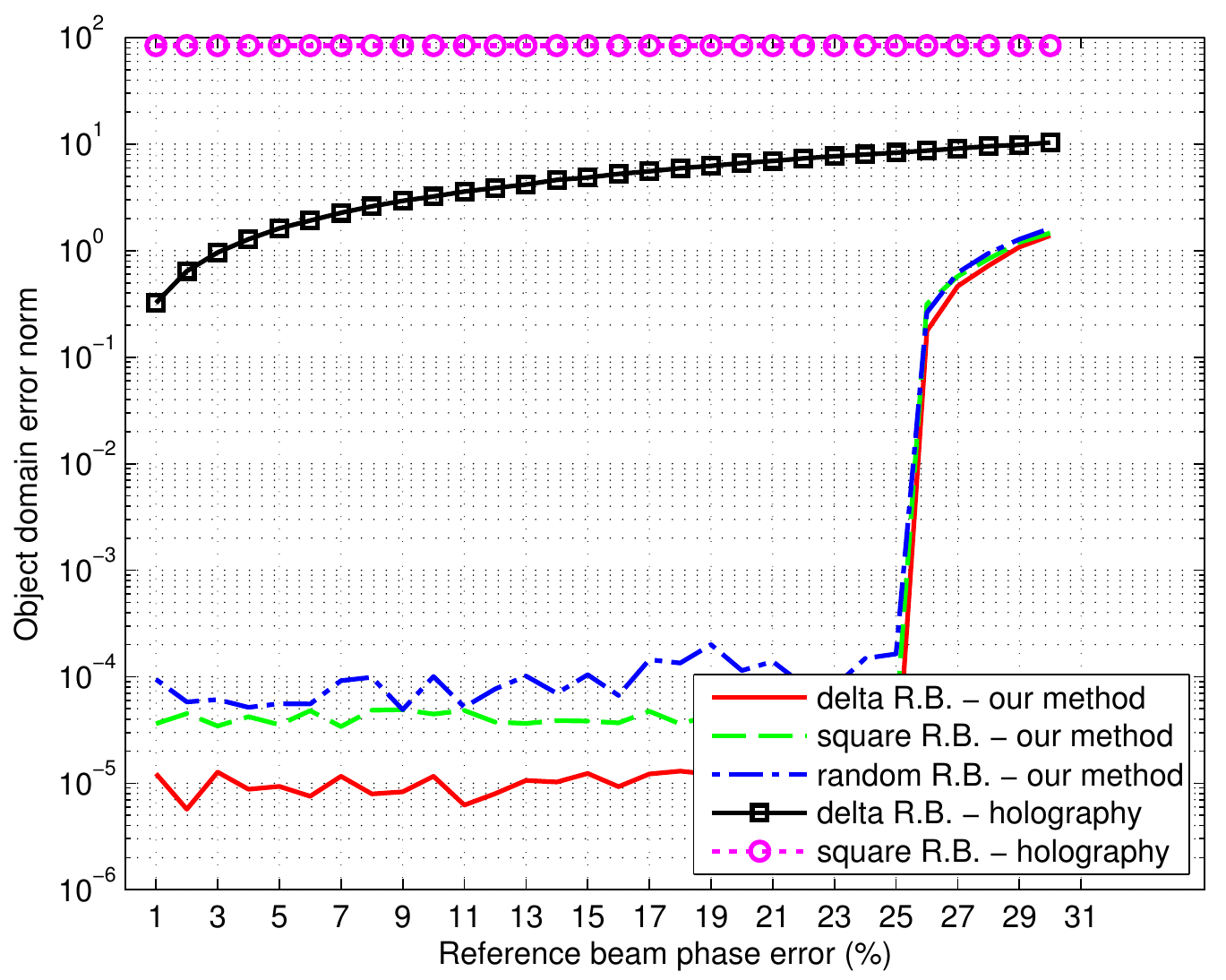}
  }
  \caption{Corrected object domain error ($\|Z-U\|$) vs. phase error in the
    reference beam: (a) real valued image,
    (b) image phase is smooth, (c) image phase is random.}
  \label{fig:objectdomain-error-corrected}
\end{figure}

\begin{figure}
  \centering
  \subfloat[]{
    \includegraphics{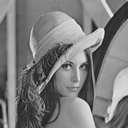}
  }
  \qquad{}
  \subfloat[]{
    \includegraphics{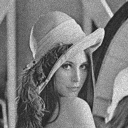}
  }
  \qquad{}
  \subfloat[]{
    \includegraphics{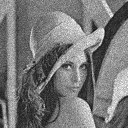}
  }\\
   \subfloat[]{
    \includegraphics{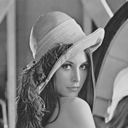}
  }
  \qquad{}
  \subfloat[]{
    \includegraphics{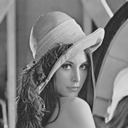}
  }
  \qquad{}
  \subfloat[]{
    \includegraphics{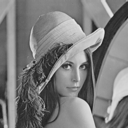}
  }
  \caption{Image (intensity) reconstructed by the holographic method
    (upper row) and our method (lower row). Object phase is random,
    and the reference beam is a delta function with Fourier phase
    errors: 
    (a), (b), and (c) --- phase error is 1\%, 10\%, and 20\% respectively
    (d), (e), and (f) --- phase error is 1\%, 10\%, and 20\% respectively.}
  \label{fig:visual-results-phase-error}
\end{figure}

In classical holography one uses a specially designed reference
beam so as to allow easy non-iterative recovery of the sought
image. This has an obvious advantage over iterative methods,
especially when the speed of the reconstruction is of high
importance. However, reliance on the reference beam means that
reconstruction quality may deteriorate badly when the reference
beam contains errors, that is, when it differs from the ``known''
values. To review the non-iterative reconstruction method used in
holography, recall that the recorded intensity $I_{2}$ is the
result of superimposition of $\hat{U}$ and $\hat{R}$ as defined by
Equation~(\ref{eq:13}). In optical Fourier holography, this
intensity is recorded on optical material. The recorded image is
used then as an amplitude modulator for an illuminating beam
$\hat{A}(\xi,\eta)$, which then undergoes a Fourier transform to
form a new signal $B(x',y')$. In a digital computer we may use the
same technique. Moreover, we are free to use either the forward or
the inverse Fourier transform, as it makes no practical difference
(the resulting images will be reversed conjugate copies of each
other). Here we use the inverse Fourier transform:
\begin{eqnarray}
  \label{eq:14}
  B(x,y)
  & = & \mathcal{F}^{-1}\{\hat{A}(\xi,\eta)\,I_{2}(\xi,\eta)\}\nonumber\\
  & = & \mathcal{F}^{-1}
  \left\{
    \hat{A}
    \left[
      |\hat{U}|^{2} + |\hat{R}|^{2}
      + \hat{U}^{*}\hat{R} + \hat{U}\hat{R}^{*}
    \right]
  \right\}\nonumber\\
  & = &
  A(x,y) \otimes U(x,y) \otimes U^{*}(-x,-y) +
  A(x,y) \otimes R(x,y) \otimes R^{*}(-x,-y) +\nonumber\\
  & &
  A(x,y) \otimes U^{*}(-x,-y) \otimes R(x,y) +
  A(x,y) \otimes U(x,y) \otimes R^{*}(-x,-y) \, ,
\end{eqnarray}
where $\otimes$ denotes convolution. Note that in this case the
fourth and the third terms are equal to the sought wavefront
$U(x,y)$ and its conjugate counterpart $U^{*}(-x,-y)$ convolved
with $A(x,y)\otimes R(x,y)$ and $A(x,y)\otimes R^{*}(-x,-y)$,
respectively.  The best possible choice is $A(x,y) = \delta(x,y)$
and $R(x,y)= \delta(x-x_{0},y-y_{0})$, where $\delta(x,y)$ is the
Dirac delta function.  In this case the obtained wave becomes:
\begin{equation}
  \label{eq:16}
  B(x,y) =
  U(x,y) \otimes U^{*}(-x,-y) + \delta(x,y) +
  U^{*}(-x+x_{0}, -y+y_{0}) + U(x-x_{0}, y-y_{0})\ .
\end{equation}
Hence, if $x_{0}$ and/or $y_{0}$ are large enough, the four terms
in the above sum will not overlap in the $(x,y)$ plane. Thus, we
can easily obtain the sought signal $U(x,y)$, albeit shifted by
$(x_{0}, y_{0})$, provided that the spatial extent of $U(x,y)$ is
limited by the box $x\in[-L_{x}, L_{y}]$, $y\in[-L_{y},L_{y}]$.
The spatial extent of the autocorrelation $U(x,y) \otimes
U^{*}(-x,-y)$ is twice as large, that is, limited by the box
$x\in[-2L_{x}, 2L_{y}]$, $y\in[-2L_{y},2L_{y}]$. Hence, to avoid
overlapping we must have $x_{0} > 3L_{x}$, or $y_{0}>3L_{y}$.
Thus, in theory, one can generate an ideal delta function in the
$(x,y)$ plane located at sufficient distance from the support area
of $U(x,y)$. In the Fourier domain, this delta function
corresponds to a plane wave arriving at a certain angle at the
plane of measurements. If such a construction is possible, then a
simple inverse transform of the intensity obtained in the Fourier
plane is sufficient to obtain the sought signal $U(x,y)$. However,
as mentioned earlier, this approach has some drawbacks. First, it
is impossible to create an ideal delta function. Any physical
realization will necessarily have a finite spatial extent, and
this will result in a ``blurred'' reconstructed image. Note that
the term ``blurring'' describes well the resulting image in the
case where $U(x,y)$ is real-valued or has constant phase. In the
more general case, where the phase of $U(x,y)$ varies at
non-negligible speed, the result appears more distorted (see
Fig.~\ref{fig:holography-reconstruction-intensity}). The second
drawback is the sensitivity of this method to errors in $\hat{R}$.
In Section~\ref{sec:results} we demonstrate how the quality of
reconstruction depends on the error in $\hat{R}$ (see
Fig.~\ref{fig:objectdomain-error}, 
\ref{fig:objectdomain-error-corrected}, and
\ref{fig:visual-results-phase-error}). Our method, on the other
hand shows very little sensitivity to the reference beam shape.
Moreover, its modification described in the next section allows
the reference beam to contain severe errors without deteriorating
significantly the quality of reconstruction.

\section{Reconstruction method for imprecise reference beam}
\label{sec:reconstr-meth-impr} Here we consider the situation
where the reference beam is not known precisely, that is, we
assume that the phase of $\hat{R}$ contains some unknown error. It
is easy to verify that if the reference beam phase
$\psi(\xi,\eta)$ has error $\epsilon(\xi,\eta)$ then the sought
phase $\phi(\xi,\eta)$ becomes
\begin{equation}
  \label{eq:15}
  \phi(\xi,\eta) = \psi(\xi,\eta) + \epsilon(\xi,\eta) \pm
  \alpha(\xi,\eta)\ ,
\end{equation}
in a manner similar to Equation~(\ref{eq:12}). We do not consider
errors in the magnitude $|\hat{R}|$ for several reasons. Many
aberrations manifest themselves through phase distortion
\cite{goodman04introduction}. Also, the magnitude of $\hat{R}$ can
be measured. Moreover, looking at the above equation, it is
obvious that any error in $\psi$ can be viewed as an error in
$\alpha$. That is, the situation would be the same if the reference
beam phase $\psi$ were known precisely, while the difference
$\alpha$ would contain some errors. This observation is relevant
because errors in the phase $\alpha$ can arise from many different
sources, including imperfect measurements and errors in the
reference beam magnitude.

The true error $\epsilon(\xi,\eta)$ is, of course, unknown. Hence,
we assume just an upper bound (assumed known) on the absolute
phase error:
\begin{equation}
  \label{eq:17}
  \psi-\epsilon-\alpha
  \leq\angle(\hat{U})\leq\psi+\epsilon+\alpha\ ,
\end{equation}
as in Equation~(\ref{eq:5}). This time, however, the phase
uncertainty interval may be larger than $\pi$ radians which makes
our method inapplicable. On the other hand, limiting the phase
uncertainty interval by $\pi$ radians will prevent us from
reconstructing the precise image, because the true phase may lie
outside this interval. A possible solution is to measure the
intensity of the reference beam and then to reconstruct its phase
using the method presented in
\cite{osherovich11approximate,osherovich09fast}, because this
problem itself can be seen as a phase retrieval with approximately
known phase. However, taking another measurement may be
undesirable, therefore, we developed the following reconstruction
method:
\begin{enumerate}
\item Set the phase uncertainty interval as defined by
  Equation~(\ref{eq:5}) (as if there were no errors in the reference
  beam phase).

\item Solve the resulting minimization problem, obtaining a
solution $Z(x,y)$).

\item If not converged, set the phase uncertainty interval to
  $[\angle(\hat{Z}) - \pi/2, \angle(\hat{Z}) + \pi/2]$. Clip it, if
  applicable, to the limits defined by Equation~(\ref{eq:17}) and go to
  Step 2.
\end{enumerate}

In this algorithm we perform a number of outer iterations, each
time updating the phase uncertainty interval. This approach leads
to a successful reconstruction method even in cases where the
reference beam contains severe errors. The results are much better
than those of non-iterative holographic reconstruction (see
Fig.~\ref{fig:objectdomain-error},
and~\ref{fig:objectdomain-error-corrected}).  This improvement is
achieved by decoupling the reconstruction problem (which becomes
the pure phase retrieval with approximately known phase) and the
erroneous interferometric measurements.

\section{Experimental results}
\label{sec:results} 
\begin{figure}
  \centering
  \includegraphics{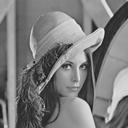}
  \caption{Original image (intensity).}
  \label{fig:experimenal-image}
\end{figure}

\begin{figure}
  \centering
  \subfloat[]{
    \label{fig:we-real-square}
    \includegraphics{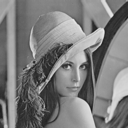}
  }
  \qquad{}
  \subfloat[]{
    \label{fig:we-complexSmooth-square}
    \includegraphics{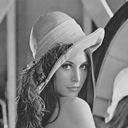}
  }
  \qquad{}
  \subfloat[]{
    \label{fig:we-complexRandom-square}
    \includegraphics{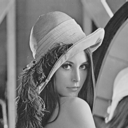}
  }\\
   \subfloat[]{
    \label{fig:we-real-random}
    \includegraphics{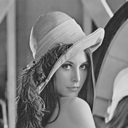}
  }
  \qquad{}
  \subfloat[]{
    \label{fig:we-complexSmooth-random}
    \includegraphics{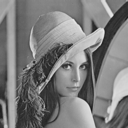}
  }
  \qquad{}
  \subfloat[]{
    \label{fig:we-complexRandom-random}
    \includegraphics{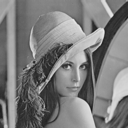}
  }
  \caption{Image reconstructed (intensity) by our method:
    (a), (b), and (c) --- reference beam is a small square and object
    phase is zero (a), smooth (b), random (c).
    (d), (e), and (f) --- reference beam is random and object phase
    is zero (d), smooth (e), random (f).}
  \label{fig:we-reconstruction-intensity}
\end{figure}

\begin{figure}
  \centering
  \subfloat[]{
    \label{fig:we-speed-real}
    \includegraphics[height=0.25\textheight]{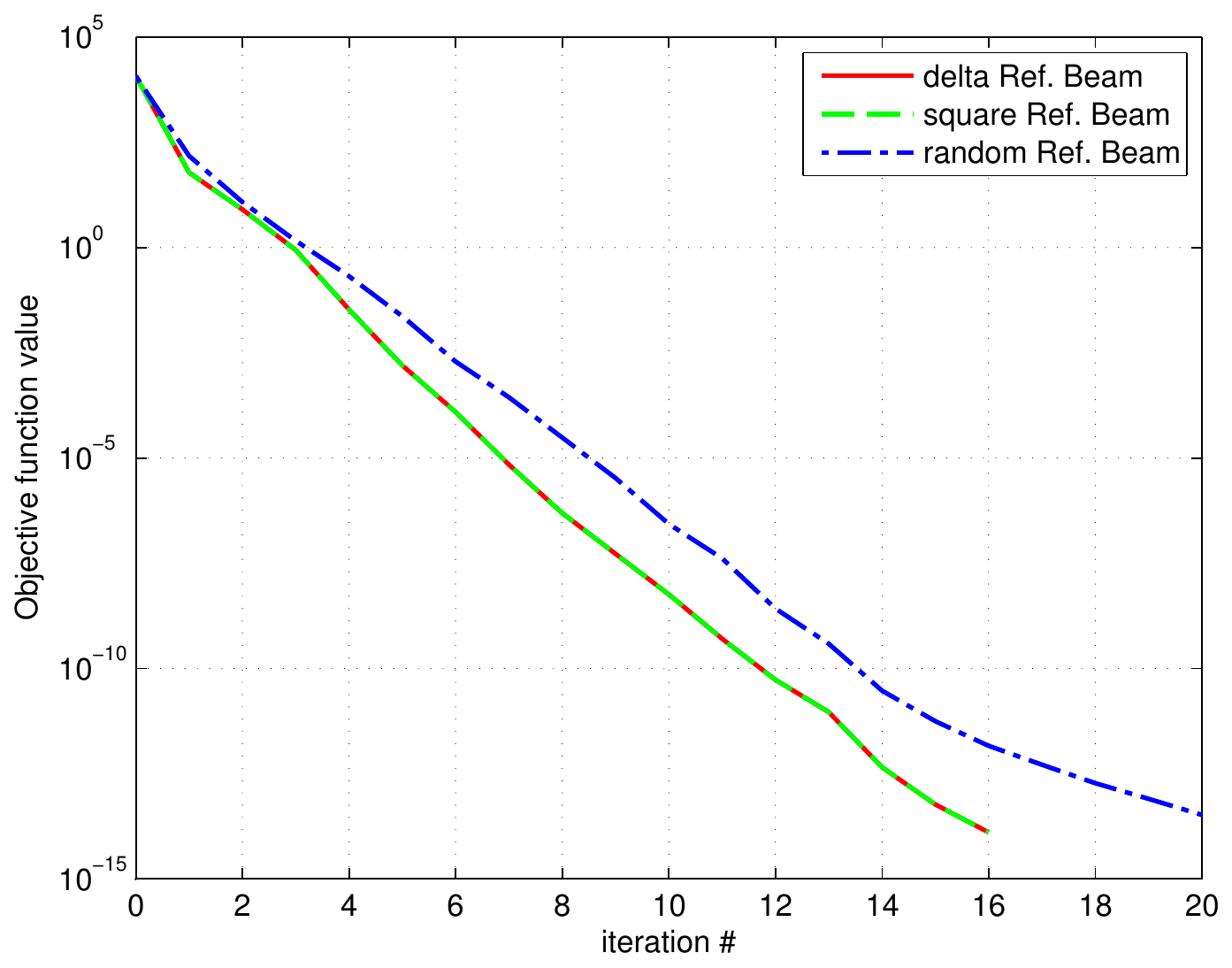}
  }
  \subfloat[]{
    \label{fig:we-speed-complexSmooth}
    \includegraphics[height=0.25\textheight]{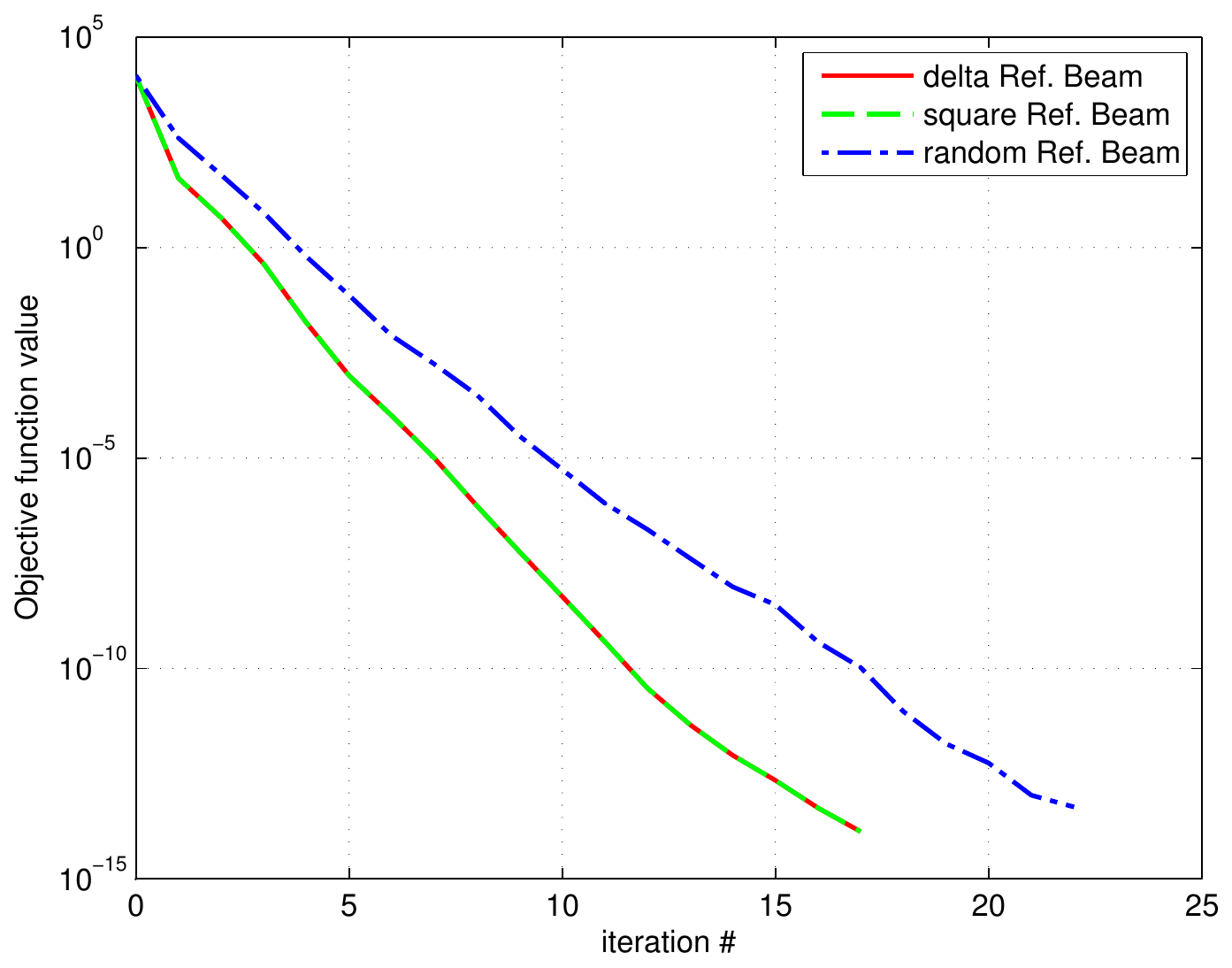}
  }
  \qquad{}
  \subfloat[]{
    \label{fig:we-speed-complexRandom}
    \includegraphics[height=0.25\textheight]{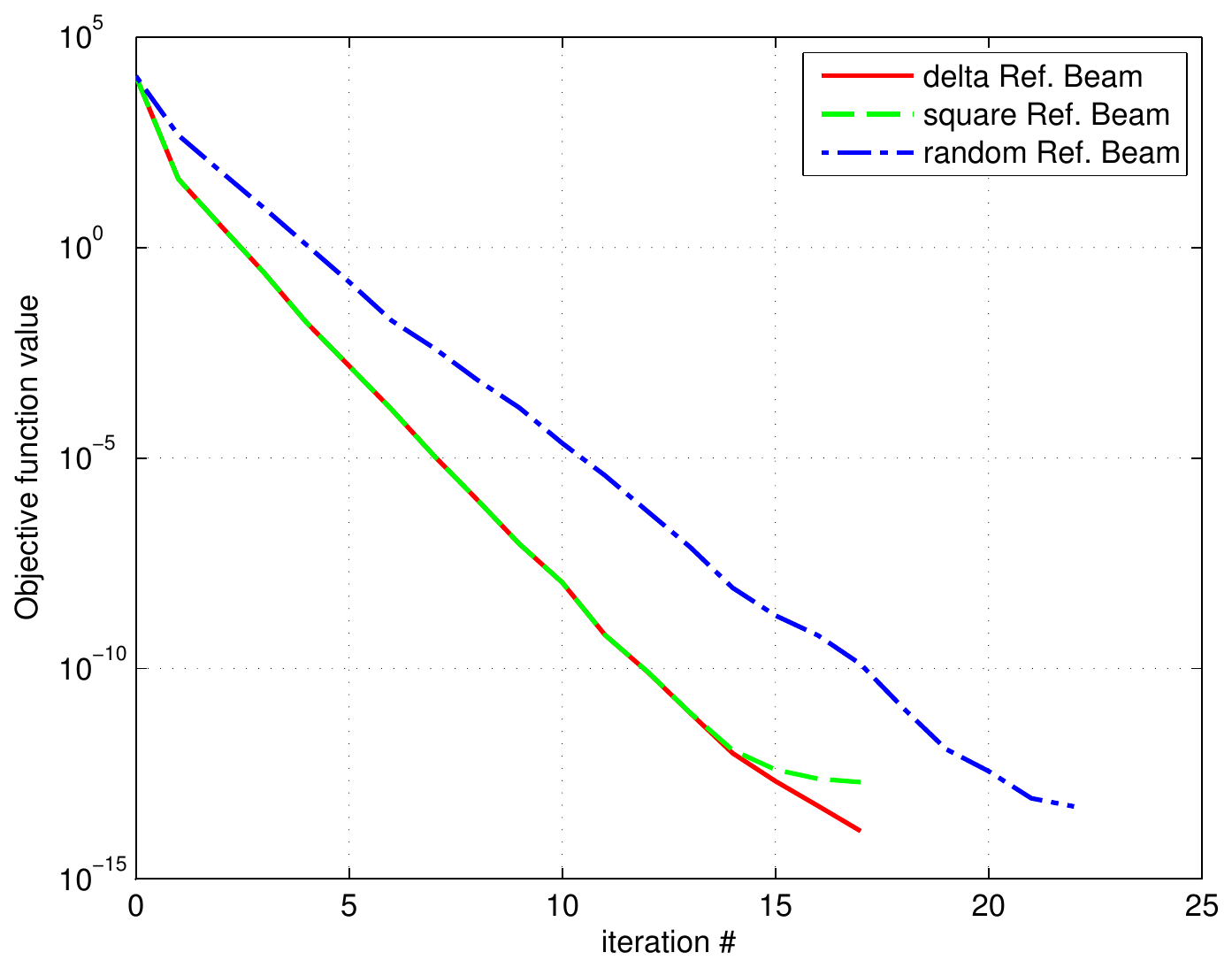}
  }
  \caption{Reconstruction speed of our method: (a) real valued image,
    (b) image phase is smooth, (c) image phase is random.}
  \label{fig:we-reconstruction-speed}
\end{figure}

The method was tested on a variety of images with similar
results. Here we present numerical experiments conducted on one
natural image so as to allow easy perception of the reconstruction
quality under various conditions. The image intensity (squared
magnitude) is shown in Fig.~\ref{fig:experimenal-image}. Technical
details of the image are as follows: the size is $128\times 128$
pixels, and pixel values (amplitude) vary from $0.2915$ to $0.9634$
with mean value of $0.6835$. These parameters will become important
when we will consider the reference beam design, and when we will
assess the reconstruction quality. Since the original image is a
photograph, it does not have any phase information. Hence, we
generated three different phase distributions to account for the
assortment of possible real-world problems where our method can be
applied. The first distribution assumes that the image is non-negative
real-valued, that is, the phase is zero everywhere. The second
distribution is designed to mimic a relatively smooth phase. To this
end, the phase is set to be proportional to the image values (scaled
to the interval $[-\pi,\pi]$). Finally, in the third distribution the
phase is chosen at random, uniformly spread over the interval
$[-\pi,\pi]$. This distribution is designed to show the behavior of
our reconstruction method in cases where the true phase varies
rapidly. We also consider three possible reference beams, again, to
demonstrate the robustness of our method. The first reference beam is
an ideal delta-function in the $(x,y)$ plane, located at the
coordinate $(256, 256)$ so that the holographic condition is
satisfied. With this reference beam exact reconstruction is obtained
as long as the sampling in the Fourier domain is sufficiently dense
($512\times512$ pixels, or more). We do not present the visual results
of reconstruction for this reference beam as both methods produce
images that are indistinguishable from the true image.  The speed of
convergence of our method is shown in
Fig.~\ref{fig:we-reconstruction-speed}.  Later, we show also how the
reconstruction quality of both methods is affected by Fourier phase
errors in the reference beam. Before this, we demonstrate the effect
of departure from the ideal delta function: the second reference beam
is a small square of size $3\times3$ pixels, located at the
coordinates $(256:258,256:258)$.  In this setup the reconstruction
quality of the holographic method is degraded, as evident from
Fig.~\ref{fig:holography-reconstruction-intensity}. It is also evident
that faster variations in the object phase result in greater
deterioration in the reconstruction, in agreement with our
expectations. Our method, on the other hand, is insensitive to the
reference beam form. In Fig.~\ref{fig:we-reconstruction-intensity} we
demonstrate our reconstruction results for the aforementioned small
square as the reference beam (the first row), and for another
reference beam that was formed in the Fourier plane by combining unit
magnitude with random phase (in the interval $[-\pi, \pi]$). This
beam, of course, is not suitable for holography, as its extent in the
object plane occupies the whole space.  Reconstruction is very fast
and, in fact, is almost independent of the sought image and reference
beam type. Fig.  \ref{fig:we-reconstruction-speed} demonstrates that
less than 20 iterations are required to solve the minimization problem
as defined by Equation~(\ref{eq:6}).

\begin{figure}
  \centering
  \subfloat[]{
    \label{fig:we-success-real}
    \includegraphics[height=0.25\textheight]{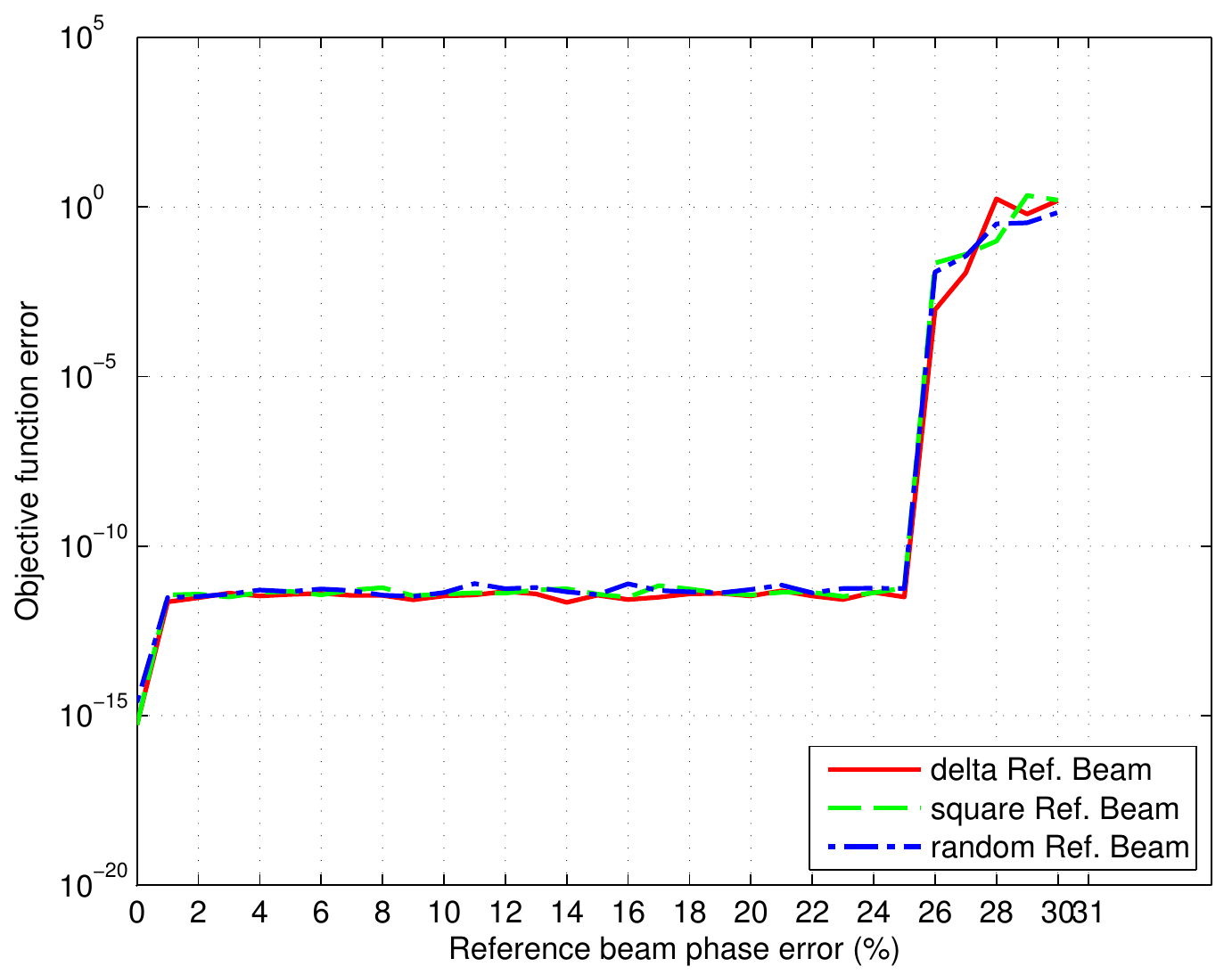}
  }
  \subfloat[]{
    \label{fig:we-success-complexSmooth}
    \includegraphics[height=0.25\textheight]{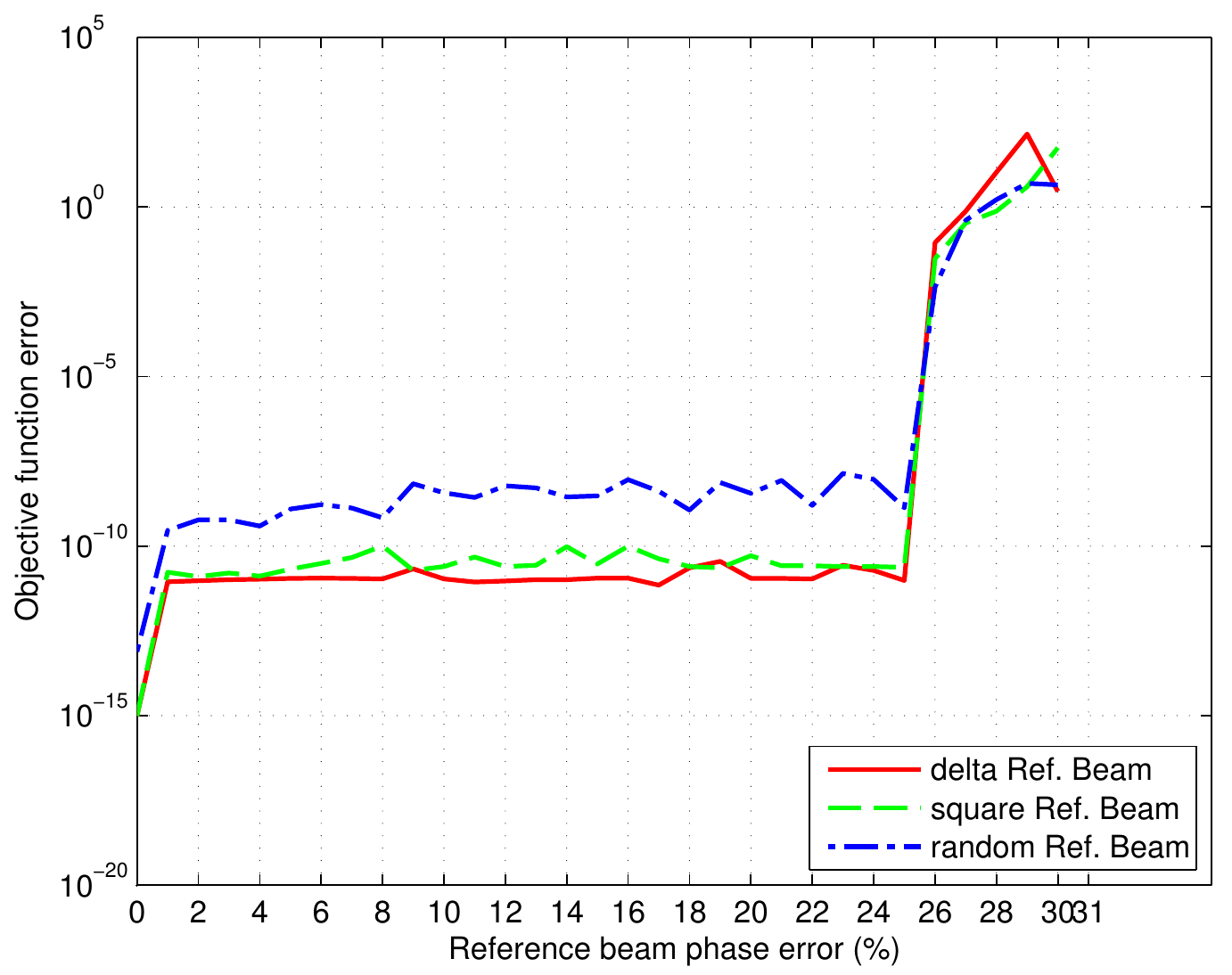}
  }
  \qquad{}
  \subfloat[]{
    \label{fig:we-success-complexRandom}
    \includegraphics[height=0.25\textheight]{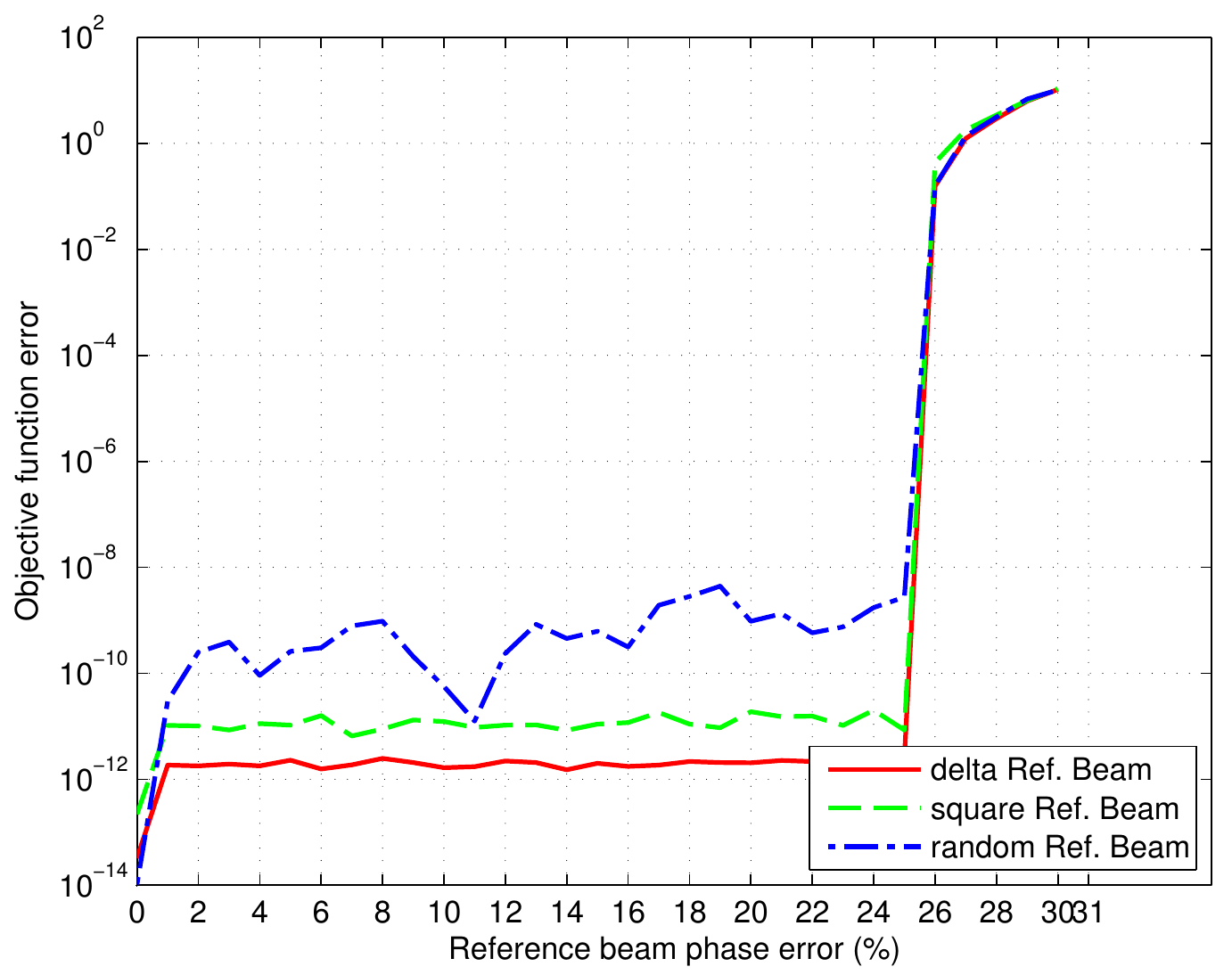}
  }
  \caption{Fourier domain ($\||\hat{Z}| - |\hat{U}|\|^{2}$) error vs. phase
    error in the reference beam: (a) real valued image,
    (b) image phase is smooth, (c) image phase is random. }
  \label{fig:we-successrate}
\end{figure}

In all these examples we assume perfect knowledge of the reference
beam. Next we consider the situation where the actual reference
beam does not match the expected signal in the Fourier plane.
Following our discussion in Section~\ref{sec:reconstr-meth-impr},
we evaluate how the reconstruction quality of the holographic
approach and our method are affected by errors in the reference
beam Fourier phase. We use again the three aforementioned models
of the sought image (real-valued, smooth phase, and random phase)
and the three reference beams (delta function, small square, and
random). From Fig.~\ref{fig:we-successrate} it is evident that in
all these cases we were able to solve the minimization problem to
sufficient accuracy as long as the phase error was below 25\%.
That is, our method can tolerate reference
beam Fourier phase errors of up to $\pi/2$ radians. The sharp
discontinuity that happens at this value has a
simple explanation: phase error greater than $\pi/2$ radians can
result in the phase uncertainty interval greater than $2\pi$ radians
(see Equation~(\ref{eq:17})).  Hence, all phase information is
lost.
A comparison
with the holographic reconstruction is given in
Fig.~\ref{fig:objectdomain-error} where the error norm in the
object domain is depicted. Note that the objective function values
are about $10^{-10}$, hence, one would expect the object domain
error norm to be of order $10^{-5}$ (the difference stems from the
fact that the objective function uses \textit{squared} norm). This
is not so in the case of complex-valued images and the random
reference beam. This effect is due to the relaxation we perform in
the Fourier phase, as discussed in
Section~\ref{sec:basic-reconstr-algor}. It does not change the
relative phase distribution, but all phases can get a constant
addition.This can be corrected by solving the one dimensional
minimization defined by Equation~(\ref{eq:19}).
Fig.~\ref{fig:objectdomain-error-corrected} depicts the corrected
values yielded by this process. Visual results comparing the
holographic reconstruction with our method are provided in
Fig.~\ref{fig:visual-results-phase-error}. 

As the results show, our method demonstrates a substantial
advantage over ordinary holographic reconstruction. It is
remarkable that even when the minimization is not particularly
successful (in cases of very large phase errors) our
reconstruction is still closer to the true signal than the
holographic method. This success is due to our approach of
decoupling the phase retrieval from the interferometric
measurements. As mentioned earlier, we deliberately avoid strong
dependence on the reference beam. The interference pattern is only
used to \textit{estimate} the Fourier phase bounds. The results
indicate that this approach is well justified.

\section{Conclusions}
\label{sec:conclusions} We present a new reconstruction method
from two intensities measured in the Fourier plane. One is the
magnitude of the sought signal's Fourier transform, and the other
is the intensity resulting from the superimposition of the
original image and an approximately known reference beam. While
the method was originally developed for the phase retrieval
problem, it can be useful in digital holography, because it poses
less stringent requirements on the reference beam. The method is
designed specifically to allow severe errors in the reference beam
without compromising the quality of reconstruction. Numerical
simulations justify our approach, exhibiting reconstruction that
is superior to that of holographic techniques.

\bibliographystyle{osajnl} \bibliography{My_Library}

\clearpage









\listoffigures

\clearpage{}

\clearpage{}

\clearpage{}

\clearpage{}

\clearpage{}

\clearpage{}

\clearpage{}

\clearpage{}

\clearpage{}

\clearpage{}

\end{document}